\documentclass[twocolumn,
showpacs,preprintnumbers]{revtex4}
\usepackage{graphicx}

\begin{document}

\preprint{YITP-03-28}

\title{BABAR resonance as a new window of hadron physics
}

\author{K. Terasaki}
\affiliation{Yukawa Institute for Theoretical Physics,
             Kyoto University, Kyoto 606-8502, Japan}

\date{May 20, 2003}

\pacs{14.40.Lb, 13.25.Ft}

\begin{abstract}
Possible decays of four-quark mesons are studied by assigning the 
newly observed BABAR resonance to a four-quark meson, 
$\hat F_I^+ \sim [cq][\bar s\bar q]$ with $q=u,\,d$. It is expected
that some of them can be observed as narrow resonances.  
Implication of existence of four-quark mesons in hadronic weak 
interactions is also discussed. 
\end{abstract}

\maketitle

Recently the BABAR Collaboration~\cite{BABAR} has observed a narrow 
$D_s^+\pi^0$ resonance with 
mass $2317.6 \pm 1.3$ MeV and width $8.8 \pm 1.1$ MeV, 
and suggested that it is a scalar four-quark meson.  

Four quark mesons, $\{qq\bar q\bar q\}$, can be classified into
four types~\cite{Jaffe}, 
\begin{equation}
\{qq\bar q\bar q\} = [qq][\bar q\bar q] \oplus (qq)(\bar q\bar q) 
\oplus \{[qq](\bar q\bar q)\pm (qq)[\bar q\bar q]\}, 
\label{eq:Jaffe}
\end{equation} 
where $()$ and $[\,]$ denote symmetry and anti-symmetry,
respectively. 
The first two on the right-hand-side of Eq.(\ref{eq:Jaffe}) can 
have $J^{P(C)}=0^{+(+)}$. Each of them is again classified into two 
classes because of two different ways to produce color singlet 
$\{qq\bar q\bar q\}$. Since these two can mix with each other, 
however, they are classified into heavier and lighter ones after all. 
We discriminate these two by putting $\ast$ on the former as in 
Table~1 according to Ref.~\cite{Jaffe} in which the four-quark mesons 
were studied within the framework of $q=u,\,d$ and $s$. To explore 
hadronic weak decays of charm mesons, we have extended the above 
framework straightforwardly to $q=u,\,d,\,s$ and $c$, and studied a 
role of four-quark mesons~\cite{charm88,charm93,charm99}. 
The heavier class of $[qq][\bar q\bar q]$ and $(qq)(\bar q\bar q)$ 
(with $\ast$) can play a dramatic role in charm decays while the 
lighter class of them (without $\ast$) can play an important role 
in hadronic weak interactions of $K$ mesons~\cite{Terasaki01}. 
Not only in hadron spectroscopy but also in hadronic weak 
interactions of $K$ and charm mesons, therefore, it is very much 
important to confirm the existence of four-quark mesons. 

In this short note, however, we will concentrate on the 
$[cq][\bar q\bar q]$ mesons (with $q=u,\,d,\,s$) since the other 
types of four-quark mesons will be massive enough to decay into two 
pseudoscalar mesons and therefore they will be broad and not very 
easy to observe. In Ref.~\cite{charm99}, however, the mass of 
${\hat F_I}\sim [cq][\bar s\bar q]$ (with $q=u,\,d$) which was 
crudely estimated by 
using a simple quark counting and the result on the light four-quark 
meson masses in Ref.~\cite{Jaffe} was higher by $\sim 100$ MeV than 
the measured one of the BABAR resonance since we took 
$\Delta m_s = m_s - m_u = 0.2$ GeV. We here revise the mass values of 
the $[cq][\bar q\bar q]$ mesons exchanging the old value of 
$\Delta m_s$ by a new one, 
$\Delta m_s \simeq m_{D_s} - m_D \simeq 0.1$ GeV 
at $\sim 2$ GeV scale, 
and using the measured $m_{\hat F_I} = 2.32$ GeV as the input data. 
[We will assign the BABAR resonance to $\hat F_I^+$ later.] 
Their revised mass values are listed in Table~1. 
[The notations have been given in Ref.~\cite{charm99}. 
$\hat E^0\sim [cs][\bar u\bar d]$ was previously 
\begin{center}
\begin{quote}
{Table~1. Ideally mixed scalar $[cq][\bar q\bar q]$ mesons 
(with $q=u,\,d,\,s)$, where 
$S$ and $I$ denote strangeness and $I$-spin.}
\end{quote}
\vspace{0.5cm}

\begin{tabular}
{|c|c|c|c|c|}
\hline
$\,\, S\,\,$
&$\,\, I=1\,\,$
&$\,\,I={1\over 2}\,\,$
&$\,\, I=0\,\,$
&$\,\,$Mass(GeV)$\,\,$
\\
\hline
$1$
&
\begin{tabular}{c}
$\hat F_I$ \\
$\hat F_I^*$
\end{tabular}
&
&
\begin{tabular}{c}
$\hat F_0$\\
$\hat F_0^*$
\end{tabular}
&
\begin{tabular}{c}
{\hspace{3mm}2.32($\dagger$)}\\
{(3.1)}
\end{tabular}
\\
\hline
0
&
&
\begin{tabular}{c}
{$\hat D$}\\
{$\hat D^*$}\\
{$\hat D^s$}\\
{$\hat D^{s*}$}
\end{tabular}
&
&\begin{tabular}{c}
{2.22}\\
{(3.0)}\\
{2.42}\\
{(3.2)}
\end{tabular}

\\
\hline 
-1
&
&
&\begin{tabular}{c}
$\hat E^0$\\
$\hat E^{0*}$
\end{tabular}
&\begin{tabular}{c}
{2.32}\\
(3.1)
\end{tabular}
\\
\hline 
\end{tabular}\vspace{2mm}\\
\hspace{-30mm}
$(\dagger)$ : Input data
\end{center}
described by 
$\hat F^0$ but it seems to be misleading so that it 
is now revised. See also, for more details, Ref.~\cite{Jaffe}.] 
To estimate more precisely the masses of $[cq][\bar q\bar q]$ 
mesons with $\ast$, we need additional input data. 

As seen in Table~1, the four-quark mesons with $\ast$ have large 
masses enough to decay into two pseudoscalar mesons so that they will 
be broad as mentioned before. On the contrary, the estimated masses 
of $[cq][\bar q\bar q]$ without $\ast$ are close to the thresholds 
of two body decays through strong interactions. Therefore, some of 
them can decay through strong interactions but their rates will be 
rather small due to their small phase space, so that they will be 
observed as narrow resonances like the BABAR resonance. 
Since some of them are not enough massive to decay into two
pseudoscalar mesons through strong interactions, their dominant 
decays may be $I$-spin non-conserving ones unless their masses are 
higher than the expected ones. 

The $\hat F_I$ mesons form an iso-triplet, $\hat F_I^{++}$, 
$\hat F_I^+$ and $\hat F_I^0$, where iso($I$)-spin symmetry is 
always assumed in this note unless we note in particular. Then all 
of them can have the same type of kinematically allowed decays, 
$\hat F_I\rightarrow D^+_s\pi$, with different charge states.  
The other decays, for example, $\hat F_I\rightarrow D_s^+(\pi\pi)$,
are not allowed kinematically.

We have two iso-doublets $\hat D$ and $\hat D^s$. The former can 
decay into $D\pi$ final states, $\hat D\rightarrow D\pi$, and the 
kinematical condition is similar to the one in the decay, 
$\hat F_I\rightarrow D^+_s\pi$, as long as the mass value of  
$\hat D$ in Table~1 is taken. The latter contains an $s\bar s$ pair 
so that its decay mode is limited because of the OZI rule~\cite{OZI}. 
One of possible decays would be 
$\hat D^s\rightarrow D\eta^s\rightarrow  D\eta$. 
Since this decay is approximately on the threshold, i.e., 
$m_{\hat D^s}\simeq m_D + m_\eta$, however, it is not clear if
such a decay is allowed kinematically, as long as the estimated value 
of $m_{\hat D^s}$ in Table~1 is taken. Even if allowed, the rate 
would be very small because of small phase space. 

$\hat F_0^+$ is an iso-singlet counterpart of $\hat F_I^+$ mesons. 
It cannot decay into $D^+_s\pi^0$ as long as the $I$-spin is 
conserved, so that it will decay dominantly through $I$-spin 
non-conserving interactions (like electromagnetic interactions). 
In this case, the width of $\hat F_0^+$ will be much narrower than 
the one of the BABAR resonance. If its mass should be higher 
(by $\gtrsim 50$ MeV) because of some $I$-spin dependent force, 
it could decay dominantly into the $DK$ final states and its width 
might reproduce the measured one of the BABAR resonance. 

$\hat E^0$ is exotic. It is an iso-singlet scalar meson with charm 
$C=1$ and strangeness $S=-1$, i.e., 
$\hat E_0^+ \sim [cs][\bar u\bar d]$. It cannot decay into $D\bar K$ 
final states unless it is enough massive because of some extra force. 
If its mass is of almost the same as the one of $\hat F_0$, it cannot 
decay through strong interactions as well as 
electromagnetic interactions~\cite{Lipkin,ST} since no ordinary 
meson with $C=1$ and $S=-1$ exists. If it can be created, therefore, 
it will be of very long life. 

Now we study numerically decays of the $[cq][\bar q\bar q]$ mesons by
assigning the BABAR resonance to $\hat F_I^+$, although there exist 
many possibilities to assign it to the other hadron states like 
a $(DK)$ molecule~\cite{BCL} (or atom~\cite{Szczepaniak}), 
an excited $(c\bar s)$  state~\cite{CJ}, an iso-singlet four-quark 
meson~\cite{CH}, etc. Consider, as an example, a decay, 
$A({\bf p})\rightarrow B({\bf p'})\, +\, \pi({\bf q})$, 
where $A$, $B$ and $\pi$ are a parent, a daughter pseudoscalar 
and a $\pi$ meson, respectively. The rate for the decay is given by 
\begin{equation}
\Gamma(A \rightarrow B\pi)
= {1\over 2J_A + 1}{q_c\over 8\pi m_A^2}
\sum_{spin}|M(A \rightarrow B\pi)|^2,
                                                    \label{eq:rate}
\end{equation} 
where $J_A$, $q_c$ and $M(A \rightarrow B\pi)$ denote the spin of $A$, 
the center-of-mass momentum of the final mesons and the decay 
amplitude, respectively. To calculate the amplitude, we here use the 
PCAC (partially conserved axial-vector current) hypothesis and a hard 
pion approximation in the infinite momentum frame (IMF), i.e., 
${\bf p}\rightarrow\infty$~\cite{suppl}. 
In this approximation, the amplitude is evaluated at a little 
unphysical point, i.e., $m_\pi^2\rightarrow 0$. 
In this way, the amplitude is given approximately by 
\begin{equation}
M(A \rightarrow B\pi) 
= \Bigl({m_A^2 - m_B^2 \over f_\pi}\Bigr)
                              \langle{B|A_{\bar \pi}|A}\rangle,
                                                     \label{eq:amp}
\end{equation} 
where $A_{\pi}$ is the axial counterpart of $I$-spin. 
{\it Asymptotic matrix elements} (matrix elements taken between single 
hadron states with infinite momentum) of $A_{\pi}$ can be 
parameterized by using {\it asymptotic flavor symmetry} (flavor
symmetry of the asymptotic matrix elements). [Asymptotic symmetry and
its fruitful results were reviewed in 
\begin{center}
\begin{quote}
{Table~2. Assumed dominant decays of scalar $[cq][\bar q\bar q]$
mesons and their estimated widths. 
$\Gamma(\hat F_I^+\rightarrow D_s^+\pi^0) = 8.8$ MeV is used as the
input data. 
The decays into the final states between $<$ and $>$ are not allowed 
kinematically as long as the parent mass values in the ( and ) are 
taken.
}

\end{quote}
\vspace{0.5cm}

\begin{tabular}
{|c|c|c|}
\hline
\begin{tabular}{c}
Parent \\
(Mass in GeV)
\end{tabular}
&Final State
&

Width 
(MeV)
\\
\hline
\begin{tabular}{c}
$\hat F_I^{++}(2.32)$ \\
$\hat F_I^+(2.32)$\\
$\hat F_I^0(2.32)$
\end{tabular}
&
\begin{tabular}{c}
$D_s^+\pi^+$\\
$D_s^+\pi^0$\\
$D_s^+\pi^-$
\end{tabular}
& 8.8 
\\
\hline
{$\hat D^+(2.22)$}
&\begin{tabular}{c}
{$D^0\pi^+$}\\
{$D^+\pi^0$}
\end{tabular}
&
\begin{tabular}{c}
{9.0}\\
{4.5}
\end{tabular}
\\
{$\hat D^0(2.22)$}
&\begin{tabular}{c}
{$D^+\pi^-$}\\
{$D^0\pi^0$}
\end{tabular}
&
\begin{tabular}{c}
9.0\\
4.5
\end{tabular}
\\
\hline 
{$\hat D^s(2.42)$}
&$D\eta$ 
&-- 
\\
\hline 
{$\hat F_0^+(2.32)$}
&\begin{tabular}{c}
{$<D_s^+\eta>$}\\
$D_s^+\pi^0$
\end{tabular}
&\begin{tabular}{c}
--\\
($I$-spin viol.)
\end{tabular}
\\
\hline 
{$\hat E^0(2.32)$}
&
$<D_s\bar K>$
&
--
\\
\hline 
\end{tabular}\vspace{2mm}\\
\end{center}
Ref.~\cite{suppl}.] Asymptotic
matrix elements including four-quark states have been 
parameterized previously~\cite{charm88,charm93,charm99}.  
We here list the related ones, 
\begin{eqnarray}
&&\langle{D_s^+|A_{\pi^-}|\hat F_I^{++}}\rangle 
= \sqrt{2}\langle{D_s^+|A_{\pi^0}|\hat F_I^{+}}\rangle 
= \langle{D_s^+|A_{\pi^+}|\hat F_I^{0}}\rangle
 \nonumber\\
&&=-\langle{D^0|A_{\pi^-}|\hat D^{+}}\rangle            
= 2\langle{D^+|A_{\pi^0}|\hat D^{+}}\rangle 
            \nonumber\\
&&=-2\langle{D^0|A_{\pi^0}|\hat D^{0}}\rangle 
= -\langle{D^+|A_{\pi^+}|\hat D^{0}}\rangle. 
                                                 \label{eq:axial-ch}
\end{eqnarray} 
Inserting Eq.(\ref{eq:amp}) with Eq.(\ref{eq:axial-ch}) into 
Eq.(\ref{eq:rate}),  we can calculate approximate rates for the 
allowed two-body decays mentioned before. Here we equate the 
calculated width for the $\hat F_I^+\rightarrow D_s^+\pi^0$ decay 
to the measured one of the BABAR resonance, i.e., 
\begin{equation}
\Gamma(\hat F_I^+\rightarrow D^+_s\pi^0) \simeq 8.8\quad {\rm MeV},  
                                                   \label{eq:input}
\end{equation}
since we do not find any other decays which can have large rates, and
use it as the input data when we estimate the rates for the other 
decays. The result is listed in Table~2. All the calculated widths of 
$\hat F_I$ and $\hat D$ are lying in the region, 4.5 -- 9.0 MeV, so
that they will be observed as narrow resonances in the 
$D_s^+\pi$ and $D\pi$ channels, respectively. 
The $\hat D^s\rightarrow D\eta$ decays are approximately on the
threshold, $m_{\hat D^s}\simeq m_D\,+\,m_\eta$, so that it is not 
clear if they are kinematically allowed. Besides, the decay is 
sensitive to the $\eta$-$\eta'$ mixing scheme which is still model 
dependent~\cite{Feldmann}. Therefore, we need more precise and
reliable values of $m_{\hat D^s}$, $\eta$-$\eta'$ mixing parameters 
and decay constants in the $\eta$-$\eta'$ system to obtain a definite 
result. 

Now we compare the decay of $\hat F_I^+\rightarrow D_s^+\pi^0$ with 
the ordinary scalar meson decay, $K_0^*(1430)\rightarrow K\pi$, 
which has been confirmed well~\cite{PDG02}. From Eq.(\ref{eq:input}), 
we can obtain $|\langle{D_s^+|A_{\pi^0}|F_I^+}\rangle| \simeq 0.12$ 
as a typical coupling strength of $\hat F_I$ to $D_s^+\pi$ using 
Eqs.(\ref{eq:rate}) and (\ref{eq:amp}). In the same way, we can 
estimate $|\langle{K^+|A_{\pi^0}|K_0^{*0}(1430)}\rangle|$ from the 
experimental data~\cite{PDG02} on the $K_0^*(1430)\rightarrow K\pi$ 
decay, i.e., from 
$\Gamma(K_0^*(1430)\rightarrow all) = 294 \pm 23$ MeV and 
$Br(K_0^*(1430)\rightarrow K\pi) = 93 \pm 10 \,\,\%$, 
we obtain $|\langle{K^+|A_{\pi^+}|K_0^{*0}(1430)}\rangle|
\simeq 0.29$. 
Since the sizes of the above matrix elements of axial charges will be 
controlled by overlappings of wavefunctions, the order of their sizes 
are quite natural, i.e., the overlapping between $q\bar q$ states (of 
$^3P_0$ and $^1S_0$) will be larger than the one between 
$[qq][\bar q\bar q]$ and $q\bar q$ (of $^1S_0$). Therefore, our 
assignment of the BABAR resonance to $\hat F_I^+$ will be again 
natural. 

In summary we have studied decays of scalar $[cq][\bar q\bar q]$ 
mesons into two pseudoscalar mesons by assigning the BABAR resonance
to $\hat F_I^+$ (in our notation) and assuming the $I$-spin 
conservation. All the allowed decays are not very far from the 
corresponding thresholds so that their rates have been expected to 
saturate approximately their total widths. Therefore, we have used 
the measured width as the input data, 
$\Gamma(\hat F_I^+\rightarrow D^+_s\pi^0) \simeq 8.8$ MeV.   
$\hat F_I$ and $\hat D$ will be observed as narrow resonances like the
BABAR one. It is very much different from the results in 
Ref.~\cite{CH} in which $\tilde D_{0s}$ [$\hat F_0^+$ in our 
notation] was assigned to the BABAR resonance and all the other 
$[c\bar qq\bar q]$ mesons were predicted to be much broader 
($\sim 100$ MeV or more). To distinguish the present assignment from
the other models and to confirm it, therefore, it is important to 
observe these narrow resonances. 

We have not studied numerically the $\hat D^s\rightarrow D\eta$. 
Since the decay is approximately on the threshold and sensitive to 
model dependent $\eta$-$\eta'$ mixing, it is hard to obtain a 
definite result on this decay at the present stage, although we can 
qualitatively expect that $\hat D^s$ will be much narrower than 
$\hat F_I$ and $\hat D$. 

$\hat E^0$ will decay through weak interactions if it is created 
as long as its mass is below the $\hat E^0\rightarrow D\bar K$
threshold. 

We have studied, so far, strong decay properties of a part of the 
four-quark mesons. If their existence is confirmed, it will be very
much helpful to understand hadronic weak decays of $K$ and charm 
mesons. The existence of $[qq][\bar q\bar q]$ and 
$(qq)(\bar q\bar q)$ mesons (with $\ast$) immediately leads to a 
solution to the long standing puzzle in charm decays~\cite{PDG02}, 
\begin{eqnarray}
&&{\Gamma(D^0\rightarrow K^+K^-)
                     \over \Gamma(D^0\rightarrow \pi^+\pi^-)}
\simeq 3,
\end{eqnarray}
in consistency with the other two-body decays of charm 
mesons~\cite{charm88,charm93,charm99}. Besides, the lighter 
$(qq)(\bar q\bar q)$ (without $\ast$) mesons are useful to 
understand the $|\Delta {\bf I}|= 1/2$ rule and its violation in 
$K\rightarrow \pi\pi$ decays in consistency with the $K_L$-$K_S$ mass 
difference, the $K_L\rightarrow \gamma\gamma$ and the Dalitz decays 
of $K_L$~\cite{Terasaki01}. 

Confirmation of the BABAR resonance as a four-quark meson will open a
new window of hadron physics.


\section*{Acknowledgments}

The author would like to thank Prof. T.~Onogi for providing
information of the BABAR resonance and encouragements. 
He also would like to appreciate  Prof. T.~Kunihiro for encouragement
and kind advises.

\end{document}